\documentclass[preprint,12pt]{elsarticle}

 \usepackage{graphicx}

\usepackage{amssymb}
\usepackage{amsmath}

\journal{Computer Physics Communications}

\begin{document}

\begin{frontmatter}


\title{FEAST fundamental framework for electronic structure calculations: Reformulation and solution of the muffin-tin problem}

 \author{Alan R. Levin, Deyin Zhang, Eric Polizzi\corref[*}
\ead{polizzi@ecs.umass.edu}
\cortext[*]{Corresponding Author}




\address{Department of Electrical and Computer Engineering, 
University of Massachusetts, Amherst, Massachusetts 01003, USA}

\begin{abstract}
In a recent article \cite{polizzi}, the FEAST algorithm has been presented as a general purpose
eigenvalue solver which is ideally suited for addressing the numerical challenges in electronic structure calculations.
Here, FEAST is presented beyond the ``black-box'' solver as a fundamental modeling framework which 
can naturally address the original numerical complexity of the electronic structure problem as formulated by Slater in 1937 \cite{slater}. 
The non-linear eigenvalue problem arising from the muffin-tin decomposition of the real-space domain is first derived 
and then  reformulated to be solved exactly within the FEAST framework.
This new framework is presented as a fundamental and practical solution for performing both accurate and scalable electronic structure
 calculations, bypassing the various issues of 
using traditional approaches such as linearization and pseudopotential techniques.
A finite element implementation of this FEAST framework along with simulation results for various molecular systems are also 
presented and discussed.

\end{abstract}


\end{frontmatter}
\section{Introduction}

Since the 1930's, progress in electronic structure calculations has always been tied together with advances in numerical strategies
 for addressing the eigenvalue Schr\"odinger equation. In particular,
several attempts have been undertaken to reduce the complexity of the eigenvalue problem in
self-consistent calculations by dissociating, screening or removing the effect of the core electrons.  
Most used techniques include \cite{martin,singhb}: muffin-tin approximations along with augmented plane wave (APW) \cite{slater}
and linearized APW, muffin-tin orbitals (MTO) and linearized MTO, KKR methods \cite{kohn}, augmented spherical wave methods \cite{eyert}, 
pseudopotential approaches \cite{hellman,pseudo} and the projector augmented-wave method \cite{blochl}. 
The conceptual approach of the former consists in partitioning the real space into spheres around each atom,   
allowing different discretization and solving strategies 
to take place in separate regions in space.
Therefore, the atom-centered regions can  benefit from specific discretization schemes (i.e. basis sets) 
that are both suitable to capture the highly localized core states around the nuclei
and considerably reduce the effective size of the resulting eigenvalue problem in the interstitial region. 
This approach can be cast as a domain decomposition method which, in modern days,  is most suitable
for parallel computing since calculations on all these sub-domains can also be performed independently.  
Once the eigenvalue problem is reformulated using domain decomposition strategies, however,
 the resulting (and still exact) problem now 
takes the form of a non-linear one in the interstitial region (i.e. $H(E)\psi=E\psi$) since the boundary conditions at the interface with the 
atom-centered regions are energy dependent. 
The major difficulty of solving this non-linear eigenvalue problem has been 
largely avoided by  mainstream approaches to electronic structure calculations 
that rely mostly on approximations ranging from direct linearization techniques
(e.g. LAPW, LMTO, etc.) \cite{andersen,singh2,sjostedt,madsen,schwarz} to pseudopotential techniques \cite{pseudo,kleinman} that eliminate
the core states.

This paper presents a fundamental strategy for performing all-electron (i.e. full-potential) electronic structure calculations 
which relies entirely on the capabilities of the new FEAST algorithm framework for solving the eigenvalue problem \cite{polizzi}. 
 At first, the algorithm can operate in parallel to obtain core and valence electrons independently spanning different
energy ranges. Secondly, solving the original eigenvalue
 problem within a given energy range (i.e. search interval) is mainly reformulated into solving a set of well-defined independent linear systems
 along a complex energy contour. 
As a result, the muffin-tin domain decomposition used to partition the real-space can act directly on the linear systems; 
therefore, the eigenvalue problem does not need to be explicitly formulated into a non-linear one. 
 In comparison to linearization techniques, the resulting linear systems also need to be evaluated for a certain set of
 “pivot energies”,  but those are now explicitly provided by FEAST to guarantee the convergence of the solutions in the entire system. 
Additionally, the complexity of interstitial problem scales linearly with the number of atoms, and it can be shown
that the proposed highly accurate all-electron muffin-tin framework is also potentially capable of better scalability performances than 
 pseudopotential approaches on parallel architectures.

The outline of this paper is as follows: In section \ref{sec:feast} we briefly summarize the numerical steps and properties of the FEAST algorithm
presented in \cite{polizzi} for solving the traditional eigenvalue problem. Section \ref{sec:muffin} presents a general definition of 
the muffin-tin strategy and the derivation of the resulting non-linear eigenvalue problem. 
Section \ref{sec:feast-muffin} describes how FEAST can be effectively and generally used to solve the muffin-tin problem without 
resorting to linearization or other approximations. The capabilities of the new all-electron framework as compared to other approaches 
are then discussed in Section \ref{sec:application} and illustrated using finite element (FEM) simulations for solving the DFT/Kohn-Sham problem 
on various molecular systems.

\section{The FEAST Algorithm}\label{sec:feast}

In electronic structure calculations, one considers solving the Schr\"odinger-type equation in an entire domain $\Omega$ which 
can be finite, periodic, or Bloch periodic:
\begin{equation}
H\Psi({\bf x})=E\Psi({\bf x}), \quad {\bf x}\in\Omega 
\label{eq_eig}
\end{equation}
where $\{E_i,\Psi_i\}$ are the resulting eigenpairs (also parametrized by $k$ in the case of bandstructure calculation using a Bloch periodic system).
Thereafter, any discretization schemes in $\Omega$ would result in the generalized and linear eigenvalue problem of size $N$:
\begin{equation}
{\bf H\Psi}=E{\bf S\Psi},
\label{eq_eigd}
\end{equation}
where $\bf S$ is a positive definite matrix (mass matrix) obtained using non-orthogonal basis functions ($\bf S=I$ otherwise), and
$\bf \Psi$ contains the $N$ unknown components of the wave function (e.g. basis set coefficients, nodal values, etc.). 

FEAST is both a new numerical algorithm \cite{polizzi} and a new general purpose high-performance numerical library \cite{feast} 
for solving the standard or generalized eigenvalue problem of type (\ref{eq_eigd}), and obtaining all the eigenvalues and eigenvectors within a given
 search interval  $[E_{min},E_{max}]$. 
FEAST is a density-matrix-based algorithm which differs fundamentally from traditional approaches for solving the eigenvalue problem (such as Lanczos, Arnoldi, Jacobi-Davidson, etc.) by combining properties of numerical linear algebra and complex analysis. 
Its main computational tasks consist of solving very few independent 
linear systems with multiple right-hand sides along a complex contour and one reduced dense eigenvalue problem orders
of magnitude smaller than the original one (the size of this reduced problem is of the order of
the number of eigenpairs inside the search interval). The basic FEAST algorithm for symmetric
eigenvalue problem detailed in \cite{polizzi} is briefly summarized in the following.

Starting from a set of $M_0$ linearly independent random vectors ${\bf {Y}}_{_{N\times M_0}}={\bf \{y_1,y_2,..y_{M_0}\}}$, 
where $M_0$ is chosen greater than the number of the eigenvalues  $M$  in the search interval
(i.e. $M_0$ represents then an over-estimation of $M$ which is not known {\em a priori}),  
a new set of vectors ${\bf {Q}}_{_{ N\times M_0}}={\bf \{q_1,q_2,..q_{M_0}\}}$ is obtained as follows:
\begin{equation}
{\bf Q}_{_{ N\times M_0}}=\displaystyle -\frac{1}{2\pi\imath}\int_{\cal C}{dZ} \ 
{\bf G}({Z}){\bf Y}_{_{N\times M_0}}, 
\label{eq_q}
\end{equation}
where ${\cal C}$ represents a complex contour from  $E_{min}$ to $E_{max}$, and the system Green's function ${\bf G}$ at the complex energy $Z$
is defined by ${{\bf G}({Z})}= ({Z}{\rm \bf S}-{\rm \bf H})^{-1}$.

\begin{figure}[htbp]
\centering
\includegraphics[width=1.0\linewidth,angle=0]{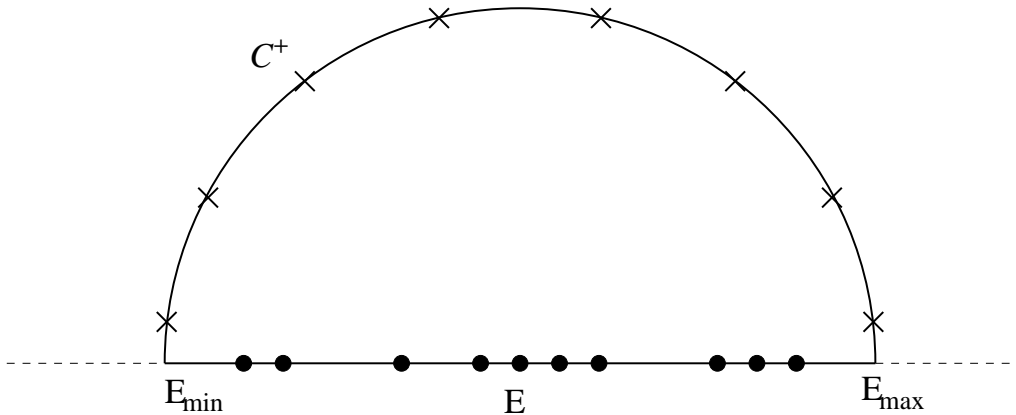}
\caption{\em \label{fig_circle} Representation of the complex contour integral defined by the positive half circle $\cal C^+$
for a given search interval $[E_{min},E_{max}]$.
In practice, the vectors $\bf Q$ are computed via a numerical integration 
(e.g. Gauss-Legendre quadrature) where only very few linear systems ${\bf G}({Z}){\bf Y}$ (\ref{eq_l}) 
needs to be solved at specific points $Z_e$ along the contour.} 
\end{figure}

In practice, the vectors $\bf Q$ in (\ref{eq_q}) can be computed using a high-order numerical integration where only very few linear systems 
${\bf G}({Z}){\bf Y}$  need to be solved along the complex contour 
${\cal C}$  i.e. 
\begin{equation}
(Z{\bf S}-{\bf H}){\bf Q^{(Z)}=Y},
\label{eq_l}
\end{equation}
where  $\bf Q^{(Z)}$  denotes the set of responses at a given pivot energy $Z$ for a given set of excitations $\bf Y$ in $\Omega$.
Since ${\bf G}({{\bar{Z}}})={\bf G^\dagger}({Z})$,
(where $\bf ^\dagger$ stands for transpose conjugate), it can be shown that the numerical integration in (\ref{eq_q}) can be performed on 
 the positive half circle of the complex contour $\cal C^+$ (see Figure \ref{fig_circle}). 

Using a Rayleigh-Ritz procedure, and by computing 
\begin{equation}
\label{eq_ral}
{\bf H_Q}_{_{M_0\times M_0}}={\bf Q^\dagger H Q} \quad \mbox{and} \quad {\bf S_Q}_{_{M_0\times M_0}}={\bf Q^\dagger S Q},
\end{equation}
a projected reduced
dense eigenvalue problem of size $M_0$ can be formed:
\begin{equation}
\label{eq_red}
\bf H_Q\Phi=\epsilon S_Q\Phi.
\end{equation}
This reduced problem can be solved using standard eigenvalue routines for dense systems  to obtain
all the eigenpairs   \mbox{$\{{\bf \epsilon}_m,{\bf \Phi}_m\}$}.
By setting  \mbox{${E}_m=\epsilon_m$} and computing \mbox{${\bf \Psi}_{_{N\times M_0}}={\bf Q}_{_{N\times M_0}}{\bf \Phi}_{_{M_0\times M_0}}$},
 it follows that if $E_m$ lies inside the contour, it is an eigenvalue solution 
 and its eigenvector is ${\bf \Psi}_m$ (the $m^{th}$ column of $\bf \Psi$). 
The eigenvectors ${\bf \Psi}$ are also naturally $\bf S$-orthonormal, 
if the eigenvectors of the reduced problem are $\bf S_Q$-orthonormal.
Accuracy can be systematically improved using a new set of initial guess vectors $\bf Y=S \Psi$  iteratively up until convergence.

The FEAST algorithm offers many important and unique capabilities for
achieving  accuracy, robustness, high-performance and scalability on parallel computing architectures. The algorithm
 holds indeed all the following intrinsic properties: 

\begin{itemize}
\itemsep 1pt
\parskip 1pt
\item Using a high-order numerical integration scheme such as Gauss-Legendre quadrature, $8$ to $16$ contour points suffice for FEAST to consistently 
converge in $\sim $2 to 3 iterations to obtain up to thousands of eigenpairs with very high accuracy. 
\item FEAST benefits from an exact mathematical derivation and 
naturally captures all multiplicities.
\item No (explicit) orthogonalization procedure is required.
\item FEAST has the ability to re-use the basis of a pre-computed subspace for fast convergence or 
as suitable initial guess for  solving a series of eigenvalue problems that are close one another (e.g. for bandstructure calculations \cite{polizzi},
 time-dependent propagation \cite{cp2010}, etc.). 
\item FEAST allows the use of iterative methods for solving the inner linear system, and which are ideally suited for large-sparse problems.
\end{itemize}

Finally, efficient parallel implementations for FEAST can be addressed at three different levels \cite{feast,fv2}:
(i) many search intervals can be run independently (no overlap),
(ii) each linear system in (\ref{eq_q}) can be solved independently  along the complex contour $\cal C$, 
(e.g. simultaneously on different compute nodes), 
(iii) the linear system (\ref{eq_l}) can be solved in parallel (the multiple right sides can be parallelized as well).
Consequently, one can show that if enough parallel computing power is available, the main computational cost 
of FEAST for solving the eigenvalue problem can be ultimately reduced to solving only one linear system (\ref{eq_l}).
This problem can be in principle addressed  by taking advantage of the many advances in ``black-box'' direct 
or iterative parallel system solvers. However, a domain decomposition strategy such as muffin-tin described in the next section,
 is naturally more suited to address specifically the electronic structure problem within a multi-atom centered environment.

\section{From Linear to non-linear eigenvalue problem using the muffin-tin strategy}\label{sec:muffin}

\begin{figure}[htbp]
\centering
\begin{minipage}{0.6\linewidth}
\includegraphics[width=\linewidth,angle=0]{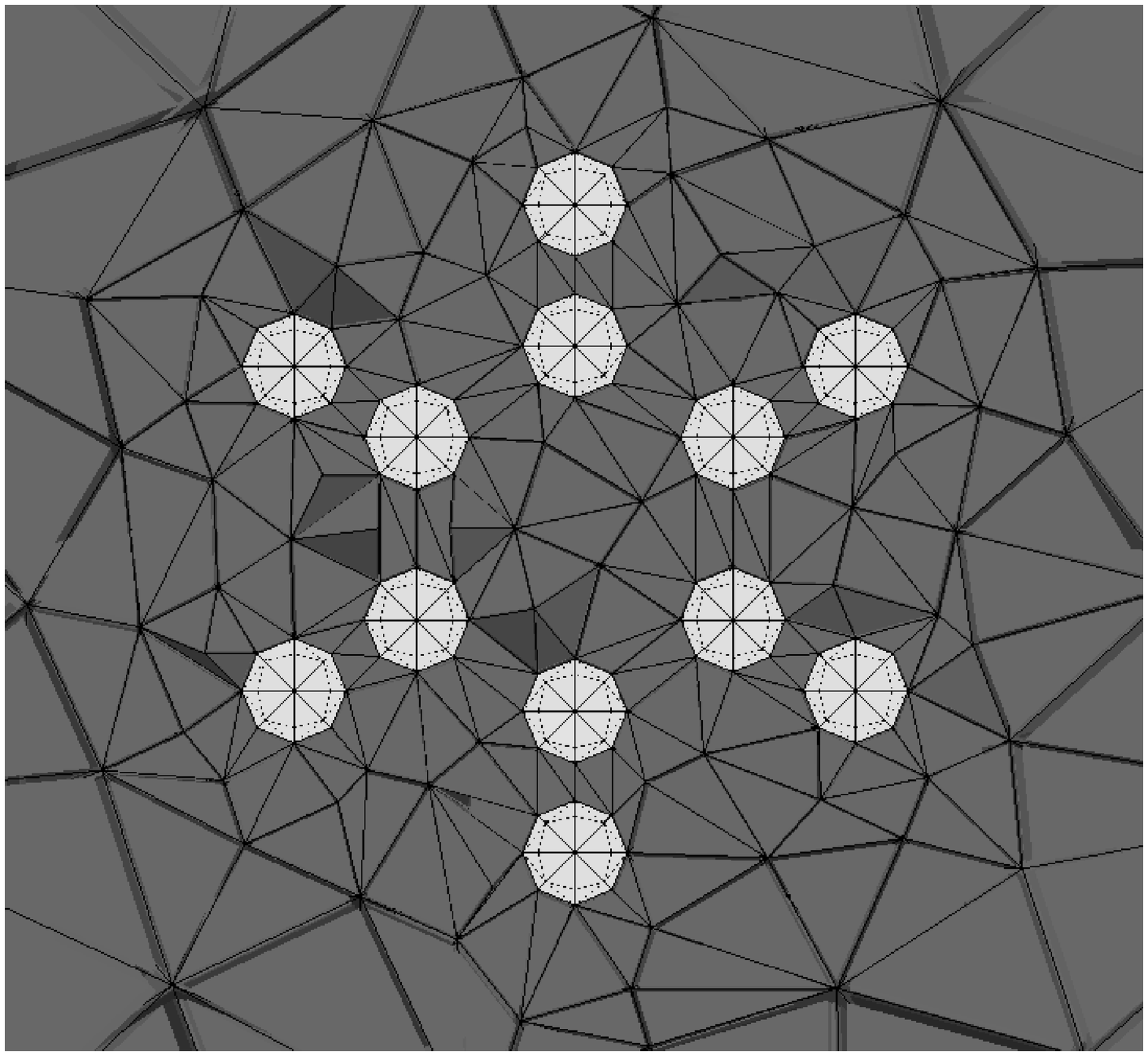} 
\end{minipage}
\begin{minipage}{0.39\linewidth}
\begin{center}
\includegraphics[width=0.6\linewidth,angle=0]{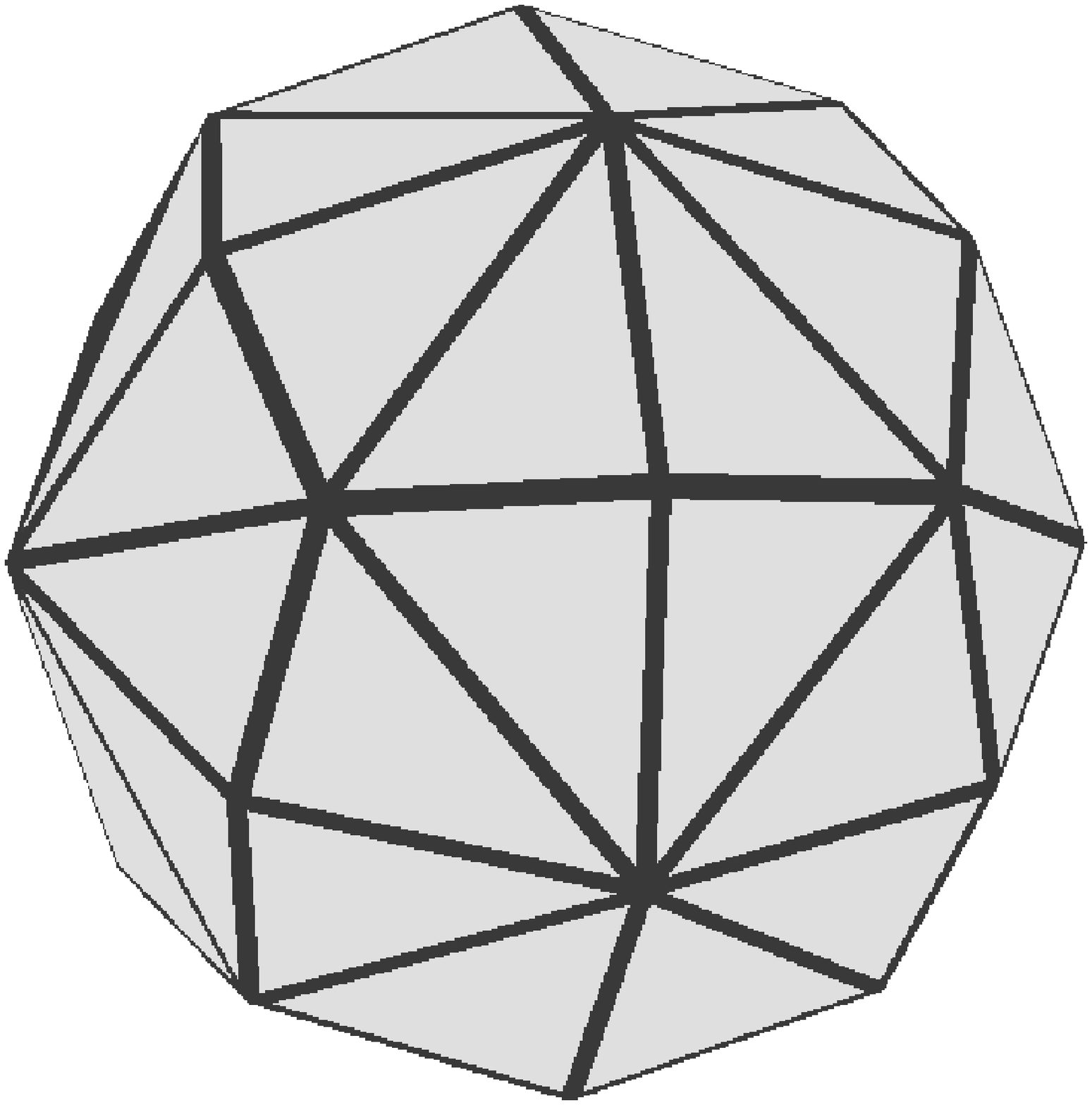}
\includegraphics[width=0.8\linewidth,angle=0]{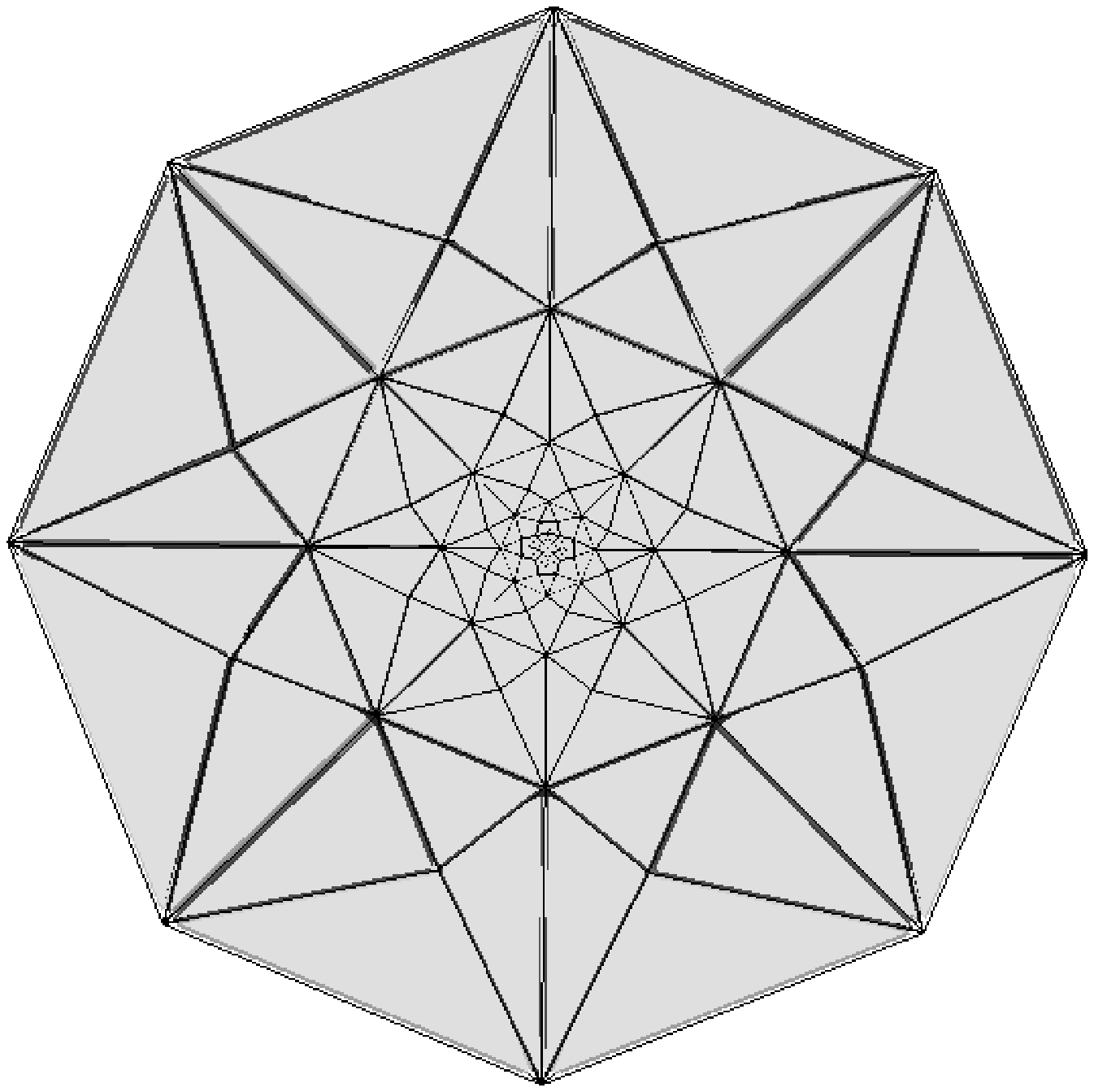}
\end{center}
\end{minipage}
\caption{\em \label{fig_benzene} 
Using a muffin-tin domain-decomposition method, the whole simulation domain $\Omega$ is separated into multiple atom-centered regions $\Omega_j$ 
($j>0$)  and one large interstitial region $\Omega_0$. Different basis-sets can be used independently to describe the different regions. 
The figure on the left represents a 2D view of local finite element discretization using a coarse mesh for $\Omega_0$ (represented only partially here) 
connecting  all of the atoms of a Benzene molecule. In contrast, the figures on the right represent a finer mesh for the $\Omega_j$  regions 
suitable to capture the highly localized core states around the nuclei.}
\end{figure}

A muffin-tin strategy for the electronic structure problem has been proposed as early as the 1930's, and in particular, 
it has been used as a starting point for the APW method introduced by Slater \cite{slater}. 
A muffin-tin decomposition brings flexibility in the discretization step, reduces the main computational efforts within the interstitial region alone, 
and should also guarantee maximum linear scalability performances using modern parallel computing architectures. 
Without any loss of generality,  Figure \ref{fig_benzene} illustrates
the essence of the muffin-tin domain decomposition (here using the particular example of a 
real-space mesh discretization for the Benzene molecule). 
For the particular case of APW, the atom-centered regions use
atomic orbitals basis functions, while a plane wave expansion scheme is used for the interstitial region. 
The derivation that follows, however, is independent of the type of basis function used within the muffin-tin decomposition.

Formally, 
the solutions $\{E_i,\Psi_i\}$ that satisfy the continuum model (\ref{eq_eig}),
 can also be obtained from a Schr\"odinger equation in the interstitial region $\Omega_0$ alone,
provided that appropriate boundary conditions are imposed at the interfaces $\Gamma_j$ with the atom-centered region $\Omega_j$ i.e.
\begin{equation}
H_0\Psi({\bf x})=E\Psi({\bf x}), \quad {\bf x}\in\Omega_0 
\label{eq_eig0}
\end{equation} 
where  $H_0$ is the Hamiltonian in $\Omega_0$. 
A general mathematical form for these boundary conditions on $\Gamma_j$ supplies a relation between the
normal derivative of the solution and their boundary values ($\forall j$):
\begin{equation}
\frac{1}{2}\frac{\partial\Psi({\bf x})}{\partial \eta_j}= \int_{\Gamma_j}  d{\bf x^{'}} \ \Sigma_j(E,{\bf x,x^{'}}) \ \Psi({\bf x^{'}}), 
\quad {\bf x}\in \Gamma_j,
\label{eq_bc1}
\end{equation}
where $\hbar=m=1$, $\eta_j$ represents the external normal at  $\Gamma_j$, and  $\Sigma_j$ is
a non-local and energy dependent operator (i.e. self-energy) which can be derived from the the atom-centered Green's function $G_j$ in 
 $\Omega_j$.  This later is given by  ($\forall j$):
\begin{equation}
\Big(E-H_j\Big)G_j(E,{\bf x},{\bf x^{'}})=\delta({\bf x}-{\bf x^{'}}), \quad {\bf x},{\bf x^{'}} \in\Omega_j,
\label{eq_Greenj}
\end{equation}
 where  $H_j$ is the Hamiltonian in $\Omega_j$, and it is important to note that
$G_j$ can be constructed with arbitrary boundary conditions at $\Gamma_j$.
For instance, by choosing the Green's function $G_j$ to have zero derivative on $\Gamma_j$ (i.e. homogeneous Neumann boundary conditions), 
one can obtain from the Green's identity a simple expression for $\Sigma_j$ (inverse of the surface Green's function):
\begin{equation}
\Sigma_j(E,{\bf x,x^{'}})=G_j^{-1}(E,{\bf x},{\bf x^{'}}) , \quad {\bf x},{\bf x^{'}} \in\Gamma_j.
\label{eq_sigma}
\end{equation}
This derivation was originally introduced in \cite{ingle} as an embedding potential technique for the Schr\"odinger equation.
Alternatively, another simple expression for $\Sigma_j$ has been derived in \cite{fisher} using 
 homogeneous Dirichlet boundary conditions for $G_j$ on $\Gamma_j$.
After discretization of (\ref{eq_eig0})  using the condition (\ref{eq_bc1}) (and usually performed on the variational 
form of the problem), the resulting non-linear eigenvalue problem in $\Omega_0$ takes the general form 
\begin{equation}
  \big({\bf H_0}-\bigcup\limits_{j}{\bf\Sigma_j}{(E)}\big){\bf \Psi_0}=E{\bf S_0\Psi_0},
\label{eq_enl0d}
\end{equation}
where $\bf S_0$ is the mass matrix in $\Omega_0$, $\bf \Psi_0$ contains the unknown components of the solution  in $\Omega_0$,
and  ${\bf \Sigma_j}(E)$ is the self-energy matrix obtained from the discretization of  (\ref{eq_sigma}) in $\Omega_j$ coupling 
 all the unknowns on $\Gamma_j$ (i.e. non-local term on $\Gamma_j$) and formally added here (for clarity) to the interstitial Hamiltonian.

Alternatively to a continuum treatment of the problem  (\ref{eq_eig}),
one could directly replace the unknown components of $\bf \Psi_j$ belonging to the interior subdomains
 $\Omega_j$ from  the system matrix  (\ref{eq_eigd}) by the following self-energy 
\cite{thij}
\begin{equation}
{\bf \Sigma_j}(E)={\bf \tau_j\ G_j \ \tau_j^\dagger},
\label{eq_fdm}
\end{equation}
 where $\bf \tau_j$ describes the interaction between $\Omega_0$ and the atom-centered region $\Omega_j$. 
In linear algebra, this non-overlapping domain decomposition procedure gives rise to a reduced non-linear system
similar to (\ref{eq_enl0d}) which is known as the Schur complement \cite{DeyinPhD}.

As originally noted by Slater \cite{slater} for the particular case of APW, this non-linear eigenvalue problem (\ref{eq_enl0d}) gives rise 
to an energy dependent secular equation which cannot be handled by traditional linear eigenvalue algorithms. Difficulties would include in particular: 
absence of orthogonality for $\bf \Psi_0$ in $\Omega_0$, and a non-linear reduced system (\ref{eq_red}) if a Rayleigh-Ritz procedure is used.
Although solving explicitly the non-linear eigenvalue problem (\ref{eq_enl0d}) is not impossible \cite{harmon,sjostedt2}, 
it remains practically very challenging.

In practice, and as  mentioned previously, the muffin-tin decomposition has always been associated with  
a new level of approximations for solving the resulting  non-linear eigenvalue problem in the interstitial region.
The mainstream approaches to all-electron electronic structure calculations  
rely indeed almost entirely on approximations such as direct linearization techniques
(e.g. LAPW, LMTO, etc.) \cite{andersen,singh2,sjostedt,madsen}. 
Alternatively, linear eigenvalue problems can also be obtained from pseudopotential techniques \cite{pseudo,kleinman} using smooth but non-local 
potentials in atom-centered regions that eliminate the core states while introducing the notion of pseudo-wavefunctions.

The next section describes how FEAST can be effectively (and generally) used to bypass these issues.

\section{Implicit treatment of the non-linear problem using FEAST}\label{sec:feast-muffin}

Here, we propose to address implicitly  the non-linear eigenvalue problem  (\ref{eq_enl0d}) by noting that
 the linear system  (\ref{eq_l}) arising from FEAST applied in the entire domain $\Omega$, can be directly solved using the same muffin-tin 
domain decomposition.

At first, starting from a set of excitations $Y({\bf x})$ in the continuum domain, 
the set of responses $Q^{(Z)}$ can also be obtained  by
solving the Schr\"odinger equation (\ref{eq_eig0}) in $\Omega_0$ alone:
\begin{equation}
(Z-H_0)Q^{(Z)}({\bf x})=Y({\bf x}), \quad {\bf x}\in\Omega_0 ,
\label{eq_eigq0}
\end{equation} 
where the boundary condition for $Q^{(Z)}$ on $\Gamma_j$ 
should formally satisfy (\ref{eq_bc1}) 
but augmented by a source term $F_j^{(Z)}({\bf x})$  (to add to the right hand side) which accounts for 
 the effects of the excitations $Y({\bf x})$ within the atom-centered regions $\Omega_j$.  
For instance, using Neumann boundary conditions for $G_j$, the self-energy $\Sigma_j$ has been defined in  (\ref{eq_sigma})
 and one can show that ($\forall j$): 
\begin{align}\label{eq_bc2}
F_j^{(Z)}({\bf x})= &  \int_{\Gamma_j}  d{\bf x'} \ G_j^{-1}(Z,{\bf x,x'})\times \nonumber \\ & \Big[ \int_{\Omega_j} d{\bf x''} \ G_j(Z,{\bf x',x''}) \ Y({\bf x''})\Big], \quad {\bf x}\in \Gamma_j.
\end{align}
Once $Q^{(Z)}$ is known in $\Omega_0$ and hence on all the $\Gamma_j$ interfaces, the solutions in the $\Omega_j$ domains 
can be independently retrieved $\forall j$ by solving the linear systems
\begin{equation}
\label{eq_ret1}
(Z-H_j)Q^{(Z)}({\bf x})=Y({\bf x}), \quad {\bf x}\in\Omega_j, 
\end{equation}
with Dirichlet boundary conditions.

After discretization of (\ref{eq_eigq0}) along with the boundary conditions (\ref{eq_bc1}), (\ref{eq_sigma}),  and (\ref{eq_bc2}), 
solving (\ref{eq_l}) in the entire domain $\Omega$ can then be replaced exactly by
the following three stage procedure:
\begin{enumerate}
\item For all $j$ atoms:
\begin{itemize}
\item obtain the self-energy ${\bf \Sigma_j}(Z)$ by computing only selected elements of the atom-centered Green's function matrix 
$\bf G_j$: 
\begin{equation}
\label{eq_gj}
{\bf G_j}(Z)={(Z{\bf {\bf S_j}-{\bf H_j})^{-1}}},
\end{equation}
\item obtain  the discretized form of the local source term ${\bf F_j}$ given by:
\begin{equation}
\label{eq_sj}
{\bf  F_j^{(Z)}}={\bf \Sigma_j}(Z) {\bf G_j}(Z){\bf Y_j},
\end{equation}
\end{itemize}

\item solve the following linear system for the unknown components of the 
solutions $\bf Q^{(Z)}_0$ in  $\Omega_0$
\begin{equation}
 \Big(Z{\bf S_0}-{\bf H_0}+\bigcup\limits_j{\bf \Sigma_j}(Z)\Big){\bf {Q}^{(Z)}_0}={\bf Y_0+\bigcup\limits_j F_j^{(Z)}}, 
\label{eq_enld}
\end{equation}
where self-energy and source term matrices in the atom-centered regions $j$ have been formally added (for clarity) to the interstitial system matrix.

\item  For all $j$ atoms, solve 
the sub-problem (\ref{eq_ret1}) to obtain the unknown components of the 
solutions $\bf Q^{(Z)}_j$ in the atom-centered regions $\Omega_j$.
\end{enumerate}
Thereafter, the subspace $\bf Q$ in (\ref{eq_q})  
is obtained by  integration of the set of solutions $\bf Q_0^{(Z)}$ and all the $\bf Q_j^{(Z)}$  over the complex contour $\cal C$.
In practice, it is possible to construct $\bf H_q$ and $\bf S_q$ in (\ref{eq_ral}) directly from the projection of $\bf H_0$ and $\bf S_0$ 
for $\Omega_0$ and  $\bf H_j$ and $\bf S_j$ for all $\Omega_j$. 

As a result of (\ref{eq_enld}) which is solved only for specific complex pivot energies $Z$,
the non-linearity of the Schr\"odinger equation (\ref{eq_enl0d}) in $\Omega_0$ is then explicitly removed, and the
muffin-tin problem benefits now from  an exact numerical treatment (i.e. no approximations needed).
In contrast to linearization techniques, these pivot energy located in the complex plane are
 explicitly provided by FEAST (e.g. Gauss contour point in Figure \ref{fig_circle}) to guarantee 
global convergence of the solutions of the Schr\"odinger equation in the whole simulation domain $\Omega$. 

In practice, it is important to mention that the additional computational costs by pivot energy $Z$ for obtaining $\bf F_j^{(Z)}$ (in step 1 above) 
and retrieving the solution $\bf Q^{(Z)}_j$ in $\Omega_j$ (in step 3 above), can be made minimal.
At first,  the muffin-tin decomposition naturally allows each atom-centered region to be factorized and solves independently.
As a result, 
the computations for obtaining  the self-energy $\bf \Sigma_j$, the source term $\bf F_j$, and 
 retrieving the solution $\bf Q_j$,  can  be fully parallelized $\forall j$. 
Additionally,  most of the 
efforts that have been devoted for obtaining ${\bf \Sigma_j}(Z)$ do not need to be repeated (e.g. factorization of the matrix 
\mbox{$(Z{\bf S_j}-{\bf H_j})$}, computations of some key elements of $\bf G_j$).


\section{Application: an all-electron real-space mesh implementation}\label{sec:application}
 In order to illustrate the efficiency and capability of the proposed FEAST muffin-tin framework, we propose to comment on our first-principle 
DFT/LDA all-electron simulations obtained on various molecular molecular systems 
 using a finite element discretization. 

\subsection{Definition of the muffin-tin finite-element mesh}

\begin{figure}[htb]
\centering
\includegraphics[width=1\linewidth,angle=0]{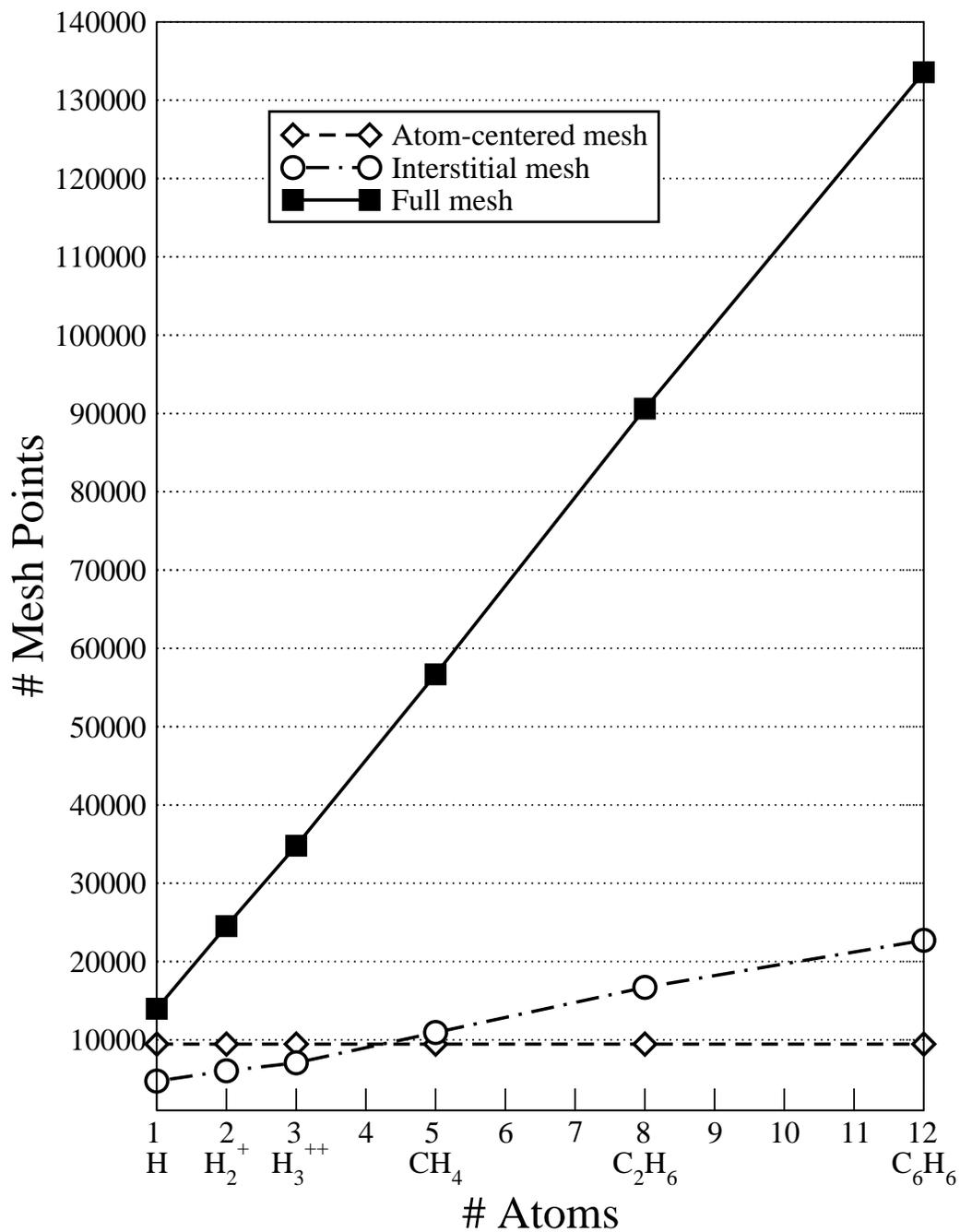}
\caption{\em \label{fig_scale}
Scalability of the number of nodes $N$, $N_j$, and $N_0$, respectively for the full, atom-centered, and interstitial meshes, with the number
of atoms per molecule. The full mesh has been reconstructed from the
muffin-tin atom-centered and interstitial meshes obtained using a cubic tetrahedra FEM discretization. 
We note here that the number of surface points at the boundary $\Gamma_j$ of the atoms is here $n_j=218$, $N_j=9457$ and 
 $N=N_0+(N_j-n_j)*N_{at}$ with $N_{at}$ the number of atoms.} 
\end{figure}

As illustrated in Figure \ref{fig_benzene}, the 3D finite-element muffin-tin mesh can be built in two steps: (i) a 3D atom-centered mesh
which is  highly refined around the nucleus, and (ii) a much coarser 3D interstitial mesh that connects all the atom-centered 
holes (both meshes have been generated using the TetGen software \cite{tetgen}).
 For the atom-centered mesh, which has been chosen common to all atoms, we have used successive layers of polyhedra similar 
to the ones proposed in \cite{puska}.
This discretization provides both tetrahedra of good quality, an arbitrary level of refinement (i.e.
 the distance between layers can be arbitrarily refined while approaching the nucleus), and the same number of 
surface nodes. Indeed, the outer layer is consistently providing the same (relatively small) number of connectivity nodes $n_j$ with
 the interstitial mesh at $\Gamma_j$ (i.e. $n_j=26$, $98$, or $218$ nodes respectively using 
linear, quadratic or cubic FEM). Consequently, 
the size of linear system (\ref{eq_enld}) in the interstitial region 
$\Omega_0$ would stay independent of  the atom-centered regions system matrix size, and the approach can then ideally deal 
with full potential (all-electron). 
Additionally and as shown in Figure \ref{fig_scale} for different molecular configurations,  the interstitial region scales linearly with the number 
of atoms while the size of the atom-centered mesh remains constant. In contrast to the (reconstructed) full mesh, the interstitial mesh
exhibits a dramatic decrease in number of mesh points and a more advantageous linear scaling rate. 

\subsection{Practical considerations}

Atom-centered, interstitial, or full finite element discretization, provide highly sparse system matrices.
As mentioned in Section \ref{sec:feast}, FEAST can be directly applied for solving the eigenvalue problem obtained from the full mesh discretization.
In particular, this can be accomplished by using the predefined sparse interfaces from the FEAST software package \cite{feast}
 (which is linked to the direct 
system solver PARDISO \cite{pardiso} by default). The software package also features a reverse communication interface that enables a straightforward 
substitution of the inner linear system factorization and solve stages along the complex contour by the muffin-tin three-step procedure 
described in Section \ref{sec:feast-muffin}.
In particular, once the linear systems (\ref{eq_eigq0}) and (\ref{eq_ret1}) formed, they can be factorized and solved using the direct system 
solver PARDISO (for example). 
We also note that only $n_j$ columns of $\bf G_j$ (\ref{eq_gj}) associated to the nodes at 
the boundary $\Gamma_j$  are needed to compute both  the self-energy ${\bf \Sigma_j}(Z)$ and the source term $\bf F_j^{(Z)}$ 
 in (\ref{eq_sj}) and this independently on the total number of nodes inside $\Omega_j$.
Indeed, after discretization of  (\ref{eq_Greenj}) and (\ref{eq_sigma}), and assuming a particular
ordering of the matrix elements (for clarity), it results
($\forall j$):   
\begin{equation}
 {\bf \hat{\Sigma_j}}(E)=\left([{\bf I}_{n_j} \ {\bf 0 \dots 0}] \ (E{\bf S_j}-{\bf H_j})^{-1} \ [{\bf I}_{n_j} \ {\bf 0 \dots 0}]^T\right)^{-1},
\label{eq_sigmad}
\end{equation}
where  the matrix $\bf \hat{\Sigma_j}$ of size $n_j$ contains all the non-zero elements of $\bf \Sigma_j$. The operation costs for (\ref{eq_sigmad})   
include solving a linear system with $n_j$ right hand sides and the inversion of a very small matrix of size $n_j$.

Additionally, one can show using a proper reordering of the matrix elements, that the same matrix factorization  
applied on ${(Z{\bf {\bf S_j}-{\bf H_j})}}$ to compute ${\bf \Sigma_j}(Z)$, can also be effectively used to compute  
 $\bf F_j^{(Z)}$, as well as the solution $\bf Q_j^{(Z)}$  (\ref{eq_ret1}) using a single solve stage \cite{AlanMs}.

\subsection{Results and Discussions on Scalability}
\begin{figure}[htbp]
\begin{center}
\includegraphics[width=0.49\linewidth,angle=0]{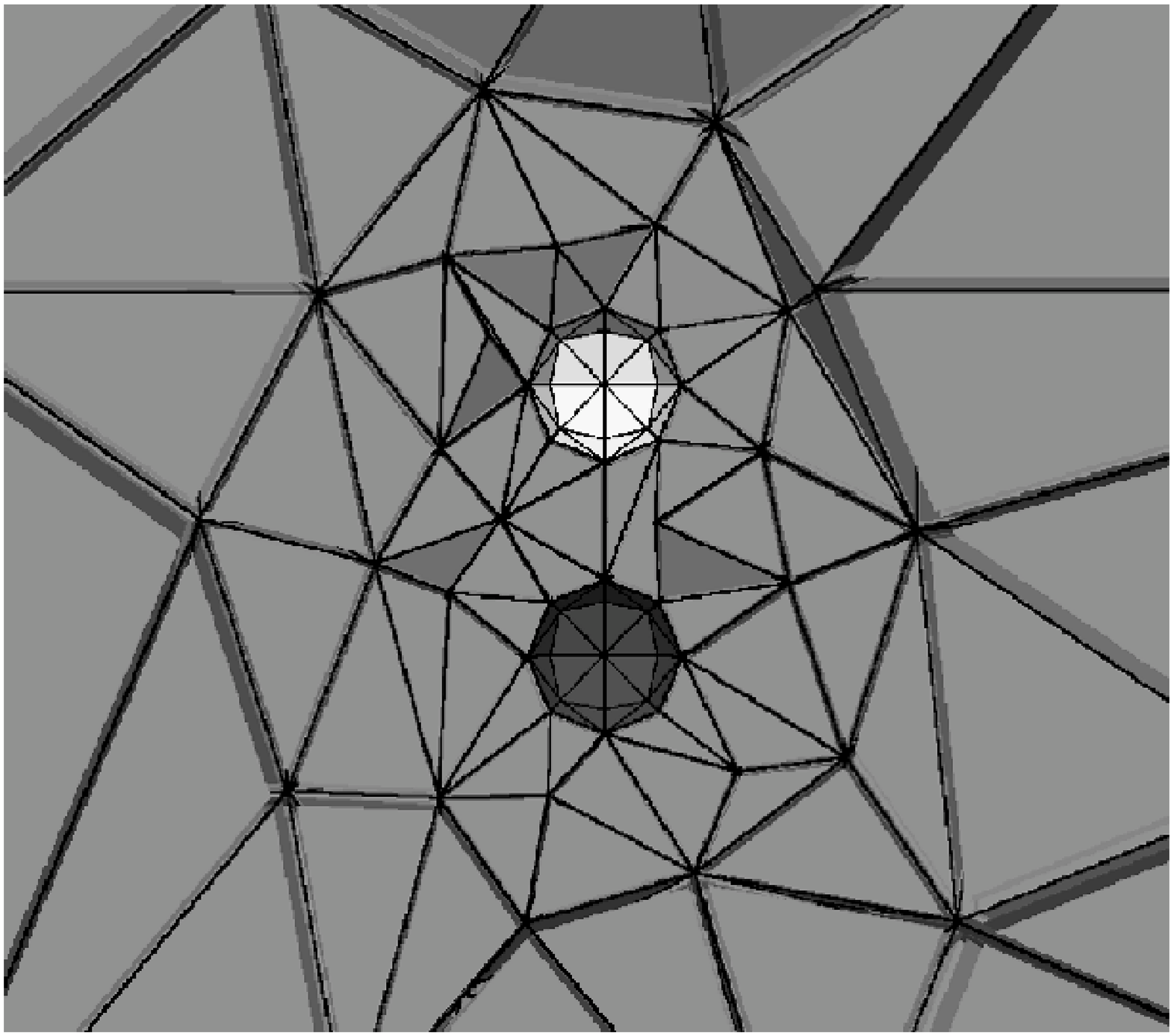}
\includegraphics[width=0.49\linewidth,angle=0]{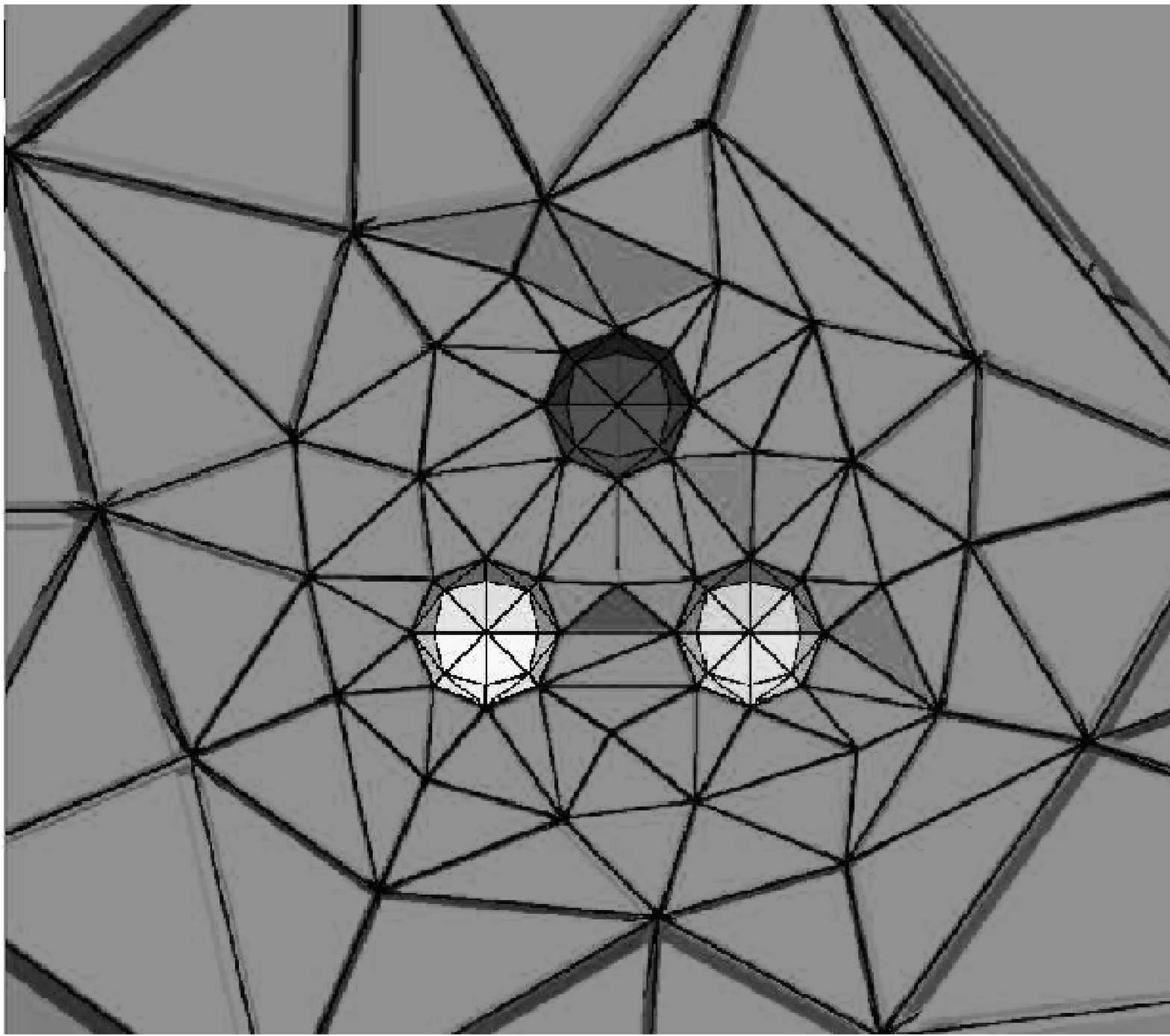}
\end{center}
\caption{\em \label{h2h3}  2D representation of the 3D
finite element discretization used to simulate the  $H_2^+$ molecule (on the left), and the $H_3^{++}$ molecule (on the right).
Only the molecular region is represented since the coarse mesh extend to $8$\AA~ (from the middle) in each direction.
All atom-centered region meshes are identical, and their radius has been set to $0.35$\AA.}
\end{figure}

At first, the results for muffin-tin FEAST framework applied to the Hydrogen atom, as well as the $H_2^+$ and $H_3^{++}$  molecules \cite{h23}, 
are compared to analytic solutions for verification purposes. For all examples, a cubic box with edges of $16$\AA~ 
was used for the interstitial region, and 
the radius of the atom-centered region has been set to $0.35$\AA~ (the proposed framework is however applicable independently of
 the choice of this atom-centered radius).
The different mesh sizes using P3 FEM have been presented in Figure \ref{fig_scale}, while a representation of the muffin-tin mesh 
for $H_2^+$ and $H_3^{++}$ 
molecules is given in Figure \ref{h2h3}.

The full-mesh and muffin-tin approaches are  fundamentally identical,
therefore the obtained numerical results  are the same (within machine accuracy). 
In comparison to analytical results, a very good agreement for the fundamental energy is obtained with $0.25\%$, $0.08\%$ and $0.05\%$ respectively 
for $H$, $H_2^+$ and $H_3^{++}$. Since the system is here bounded by a fixed size simulation box, 
by increasing the number of atoms (or atomic weight), one also increases the locality of the solutions in the molecular region 
and the overall accuracy.
In a more general situation, accuracy can be systematically increased by refining the atom-centered mesh along with
 the interstitial mesh in the molecular region (i.e. increasing the number of nodes or the order of the FEM).

For the case of the Benzene molecule (see Figure \ref{fig_benzene}), all-electron simulations have been performed self-consistently
using both a GR-pulay procedure \cite{pulay}, and an alternative approach deriving from a generalization of the 
FEAST algorithm for solving directly the full coupled non-linear DFT/Kohn-Sham problem \cite{bp}.
While a full three-dimensional discretization using cubic finite element gives rise to sparse system matrices of size 
of $N=133,579$ in $\Omega$ (see Figure \ref{fig_scale}),  the muffin-tin decomposition provides, in turn, a series of subsystems of size
  $N_j=9457$  in the atom-centered regions $\Omega_j$ with $n_j=218$ on $\Gamma_j$, and a single system of size $N_0=22,711$ in the interstitial 
region $\Omega_0$.  Our total energy found at self-consistent convergence 
 $E_{\rm LDA}=-6261.14$ eV, is also in very good agreement with the accuracy of the all-electron simulation results presented in \cite{puska}.

The muffin-tin decomposition naturally allows each atom-centered
 region to be factorized and solved independently; 
as a result, 
the computation for obtaining the self-energy $\bf \Sigma_j$, the source term $\bf F_j$, and 
 retrieving the solution $\bf Q_j$,  can be fully parallelized. 
Since the scalability of FEAST on parallel architecture is first and foremost dependent on the
scalability of the inner linear system solver (i.e. third level of parallelism for FEAST), the main difficulty would come from solving 
the interstitial system (\ref{eq_enld}) in parallel.
As shown in Figure \ref{fig_scale}, the size of the interstitial system  does increase linearly but
at a much lower rate than the full system (i.e. $N_0<<N$ for large number of atoms).
For example, while $N=133,579$ is used for the full twelve atoms system of the Benzene molecule, the same
size order would only be reached by the interstitial system ($N_0$) using $85$ atoms. Similarly, $500$ atoms would only generate an interstitial  
system of moderate size $N_0=750,000$, while $N\simeq 5\times10^6$ would be required for the full system (using $N_j\simeq 10000$).
It is important to note that the muffin-tin approach enables the use of hybrid basis-set, and  a quadratic rather a cubic FEM discretization could 
then potentially be used for the interstitial region that exhibits low-varying potential (and $N_0$ would then decrease significantly). 
Finally,   the size $n_j$ of the non-local blocks in $\bf \Sigma_j$ (\ref{eq_sigmad}) being
 relatively very small, 
the linear scalability of the linear system (\ref{eq_enld}) is then not 
 affected by the presence of the self-energy matrices.  
Consequently, in practice the interstitial system can be efficiently solved using  
``black-box'' parallel sparse system solver that can be either direct or iterative (using an appropriate preconditioner).

\subsection{FEAST all-electron calculations vs pseudopotential}

The FEAST framework is independent of any particular form of the potential in the atom-centered regions (such
as spherically symmetric potential, etc.). Since  the size of linear system (\ref{eq_enld}) in the interstitial region 
$\Omega_0$ is independent of the discretization schemes for the atom-centered regions $\Omega_j$, the approach can ideally deal with full potential 
 within self-consistent calculations (i.e. all-electron calculations).

The development of techniques such as pseudopotential were originally motivated to ease 
several numerical difficulties that one can encounter with all-electron calculations in the atom-centered regions \cite{hellman,pseudo,kleinman}.
Let us then outline how some of these main issues are naturally addressed within the FEAST all-electron framework:

(i) Since FEAST can act independently on different energy ranges (with no overlap), the number of states in a search interval can be narrowed as desired, 
and the frozen-core approximation does not need to be considered within self-consistent iterations. 

(ii) In contrast to pseudopotential, it is clear that a  much finer level of discretization 
for the FEAST all-electron framework is needed to capture 
the (true) wave functions in the atom-centered regions $\Omega_j$ (i.e. the pseudo-wavefunctions can be captured within a reduced basis set).
The linear eigenvalue system obtained using pseudopotential can either be seen as a much smaller size system as compared to (\ref{eq_eigd}), 
or a linearized version of (\ref{eq_enl0d}) where $\bf \Sigma_j$ represents then a fully non-local pseudopotential over the atom-centered 
region $\Omega_j$. Although this pseudopotential system can also ideally be solved using the FEAST algorithm, 
the resulting system matrix (\ref{eq_l}) 
would end up (paradoxically) being larger and less scalable than (\ref{eq_enld}) (which only requires the surface terms $\Gamma_j$ of 
the $\Omega_j$ regions).

\section{Conclusion}

In 1937, Slater originally derived a non-linear electronic structure problem by introducing the APW method using 
a muffin-tin domain decomposition; he then stated \cite{slater},
{\it ``Of course, we cannot solve this exactly, and we must look for methods of approximations''}. 
These limitations have historically motivated the development of a wide spectrum of approximation techniques ranging from direct linearization to
pseudopotential methods.
Within the framework of the FEAST algorithm,  however, the muffin-tin problem benefits now from an exact numerical treatment (i.e. no approximation needed) 
 which consists of removing the non-linearity of the Schr\"odinger equation (\ref{eq_enl0d}) in the interstitial region $\Omega_0$, 
by considering only certain complex pivot energies $Z$ (\ref{eq_enld}). 
In contrast to linear approximations  (including LAPW, LMTO, 
linearized embedding method \cite{ingle}, etc.), these pivot energies are  explicitly provided by FEAST with guaranteed 
convergence of the solutions of the Schr\"odinger equation in 
the whole simulation domain. 
As compared to ab-initio pseudopotential approaches, in fact, not only the proposed  FEAST framework 
is more accurate by essence (since all-electron calculations are performed), it is also capable of higher parallel scalability performances.
In conclusion, the new approach is not tied together with the traditional modeling
trades-off between robustness/accuracy and performances/scalability in simulations. 

Finally, the FEAST fundamental framework for first-principle electronic structure calculations can be used
independently of the choice for the physical model (e.g. Density-Functional Theory or Hartree-Fock), the nature of the atomistic system 
(e.g. isolated or Bloch periodic), or the choice for the basis set (e.g. PW, atomic orbitals, real-space mesh).

\section*{Acknowledgments}
This material is based upon work supported by the National Science Foundation under Grant No ECCS 0846457.

\bibliographystyle{elsarticle-num}
\bibliography{lzp11-cpc}


\end{document}